\documentclass[manuscript,screen,sigconf,nonacm,natbib=false]{acmart}
\renewcommand
\footnotetextcopyrightpermission[1]{} % removes footnote with conference information in first column
\makeatletter
\renewcommand\@formatdoi[1]{\ignorespaces}
\makeatother

\AtBeginDocument{%
  \providecommand\BibTeX{{%
    \normalfont B\kern-0.5em{\scshape i\kern-0.25em b}\kern-0.8em\TeX}}}

\usepackage{caption}
\usepackage{subcaption}
\usepackage{makecell}
\usepackage[style=ACM-Reference-Format,backend=bibtex,sorting=none]{biblatex}
\setcopyright{none}
\begin{document}

%%
%% The "title" command has an optional parameter,
%% allowing the author to define a "short title" to be used in page headers.
\title{Evaluating Transferability for Covid 3D Localization Using CT SARS-CoV-2 segmentation models}

%
% The "author" command and its associated commands are used to define
% the authors and their affiliations.
% Of note is the shared affiliation of the first two authors, and the
% "authornote" and "authornotemark" commands
% used to denote shared contribution to the research.
\author{Constantine Maganaris}
\affiliation{
  \institution{National Technical University of Athens, School of Rural, Surveying and Geoinformatics Engineering}
  \city{Athens}
  \country{Greece}}
\email{c@maganaris.com}
\authornote{All authors contributed equally to this research.}
\orcid{0000-0003-2738-684X}
\author{Eftychios Protopapadakis}
\authornotemark[1]

\affiliation{%
  \institution{National Technical University of Athens, School of Rural, Surveying and Geoinformatics Engineering}
%   \streetaddress{P.O. Box 1212}
  \city{Athens}
  \country{Greece}
%   \postcode{43017-6221}
} 
\email{eftprot@mail.ntua.gr}

\author{Nikolaos Bakalos}
\affiliation{%
  \institution{Institute of Communications and Computer Systems}
%   \streetaddress{1 Th{\o}rv{\"a}ld Circle}
  \city{Athens}
  \country{Greece}}
  \email{bakalosnik@mail.ntua.gr}

 \author{Nikolaos Doulamis}
\affiliation{%
  \institution{Institute of Communications and Computer Systems}
%   \streetaddress{1 Th{\o}rv{\"a}ld Circle}
  \city{Athens}
  \country{Greece}}
\email{ndoulam@cs.ntua.gr}

\author{Dimitris Kalogeras}
\affiliation{%
    \institution{Institute of Communications and Computer Systems}
%   \streetaddress{1 Th{\o}rv{\"a}ld Circle}
  \city{Athens}
  \country{Greece}}
\email{d.kalogeras@noc.ntua.gr}

\author{Aikaterini Angeli}
\affiliation{%
    \institution{Institute of Communications and Computer Systems}
%   \streetaddress{1 Th{\o}rv{\"a}ld Circle}
  \city{Athens}
  \country{Greece}}
\email{katerina.angeli.doulami@gmail.com}

%%
%% By default, the full list of authors will be used in the page
%% headers. Often, this list is too long, and will overlap
%% other information printed in the page headers. This command allows
%% the author to define a more concise list
%% of authors' names for this purpose.
% \renewcommand{\shortauthors}{XXXXXXXXXXXXXXXXXXXXXXXXXXXX}
\renewcommand{\shortauthors}{C. Maganaris, E. Protopapadakis, N. Bakalos, N. Doulamis, D. Kalogeras and A. Angeli}

%%
%% The abstract is a short summary of the work to be presented in the
%% article.
\begin{abstract}
  Recent studies indicate that detecting radiographic patterns on CT scans can yield high sensitivity and specificity for Covid-19 localization. In this paper, we investigate the appropriateness of deep learning models transferability, for semantic segmentation of pneumonia-infected areas in CT images. Transfer learning allows for the fast initialization/reutilization of detection models, given that large volumes of training data are not available. Our work explores the efficacy of using pre-trained U-Net architectures, on a specific CT data set, for identifying Covid-19 side-effects over images from different datasets. Experimental results indicate improvement in the segmentation accuracy of identifying Covid-19 infected regions.
\end{abstract}

%%
%% The code below is generated by the tool at http://dl.acm.org/ccs.cfm.
%% Please copy and paste the code instead of the example below.
%%


%%
%% Keywords. The author(s) should pick words that accurately describe
%% the work being presented. Separate the keywords with commas.

%%
%% This command processes the author and affiliation and title
%% information and builds the first part of the formatted document.
\maketitle
\thispagestyle{empty}

\section{Introduction}
Coronavirus pandemic (COVID-19) and its variations, has infected more than 427 million people and caused more than 5,9 million deaths around the globe, based on the facts of John Hopkins University (22 February 2022) \cite{noauthor_Covid-19_nodate}. Since the declaration of Public Health emergency of International Concern on January 30 2020 \cite{noauthor_statement_nodate} from the World Health Organization (WHO) there was a research outbreak in the field, to respond the emergency situation. A great amount of research has been focused since then on fast detection of COVID, using CT or X-rays of the thorax, in parallel with other methods like Antigenic and Reverse Transcription-Polymerase Chain Reaction (RT-PCR) testing. 

Currently, the WHO recommends the use of rapid tests \cite{noauthor_antigen-detection_nodate} in general population for primary case detection in symptomatic individuals suspected to be infected and asymptomatic individuals at high risk of COVID-19, for contact tracing, during outbreak investigations and to monitor trends of disease incidence in communities, even if they have poor sensitivity in their results \cite{ferte_accuracy_2021}. The RT-PCR tests are also known to have relatively high false negative rates \cite{kortela_real-life_2021}. This made crucial to invent methodologies to detect fast and reliably COVID infection but also to estimate the cruciality of it or how fast it expands in a patient, to curate it better, these data are not available with the methods of PCR or Antigen testing. This makes the use of CT and X-ray scans a mandatory tool in the hands of clinicians to work with COVID patients, but also suggests the need for creating new tools using Artificial Intelligence to assist the experts in their work and make detection of the effects of COVID faster. 

It was 2016, at a Conference in Toronto, when Geoffrey Hinton said that “[...] People should stop training radiologists now, it is just completely obvious that in 5 years Deep learning is going to do better than radiologists because this can be able to hit a lot more experience. It might be 10 years but we got plenty of radiologists already” \cite{creative_destruction_lab_geoff_2016}. Nevertheless, things progressed differently. During the pandemic, it became clear that the AI tools cannot replace radiologists, which were much in need and the pandemic had a great impact in them, especially in cases like Northern Italy \cite{coppola_impact_2021}. 

Despite all that, it is crucial to continue developing automated decision making tools to assist healthcare personnel and overcome all the issues that comes with the analysis of Computed Tomography Imaging. One of the main setbacks in creating accurate models is the lack of publicly available datasets with chest scans of infected people. In addition, a lot of data that have been used so far, have faulty cleansing, so they create frustration on the results produced by deep learning models, trained to detect COVID \cite{noauthor_hundreds_nodate}. Moreover, scans coming from different equipment, produce data that differentiate on how they signal COVID areas using the Hounsfield scale thus making the training of appropriate models even more difficult due to the significant differences in the training data.

In our research we use two different datasets in order to detect the transferability of deep learning models between them, and estimate the accuracy of them to detect segments of COVID areas (i.e. markings of ground glass opacities, consolidation and pleural) \cite{cozzi_ground-glass_2021}. These datasets include 3410 slides of CT scans with annotated COVID and Lung segments, but also have masks of the aforementioned signals of COVID-19.

From our results we have found that there is evidence of such transferability of results, after retraining the model in a portion of the data of a second dataset and for a short number of epochs for the second train. It is important to note that we achieved a great prediction outcome with good rankings in precision, recall and F1 metrics. 
To present our results, we have visualized 3D reconstructions of the lungs using data from the CT scans. This way we achieved better understanding on how the model works in the complete CT scan and not by comparing separate slides of it. In our 3D model reproduction, we used only the lung areas of the slides, the annotations of the COVID areas made by the radiologists (ground truth) and the predicted COVID areas from our model. Now having these representations in hand we can view the bigger picture and how the model performs in a patient. This is making it easier for a user (i.e. a doctor) of the model to extract information from the results of how serious the illness is, or how the infection responds to medication etc. by comparing different scans of the same patience in different points of time. 

Another finding of our research was that even in our set of data, there were COVID areas (GGOs) falsely not annotated by the radiologists, we have seen that our model managed to correctly annotate COVID affected areas. Additionally, areas that have been falsely marked as COVID by the radiologists, and the model has been trained in them, produces correct output (i.e. non Covid) from our model.

\section{Related work}
Deep learning methodologies using various types of images are common for identfication, detection or segmentation in medical imaging \cite{voulodimos_deep_2018} and in biomedical applications \cite{le_fertility-gru_2019}. In this context, researchers already investigating several approaches to assist medical professionals with Covid-19 detection.
An initial approach was to classify multiple CT slices using a convolutional neural network variation \cite{li_coronavirus_2020}. The adopted methodology is able to identify a viral infection with a ROCAUC score of 0.95 (a score of 1 indicates a perfect classifier). However, despite the high detection rates, the authors indicated that it was extremely difficult to distinguish among different types of viral pneumonia based solely on CT analysis.

Convolutional Neural Network (CNN) variations for the distinction of coronavirus vs. non-coronavirus cases have been proposed by \cite{li_using_2020}. The specific approach allows for a distinction among Covid-19, other types of viral infections, and non-infection cases. Results indicate that there are adequate detection rates and a higher detection rate than RT-PCR testing. Towards this direction, CNN structures are combined with Long Short Term Memory (LSTM) networks to improve the classffication accuracy of CNN networks further \cite{islam_combined_2020}. Additionally the work of \cite{fan_inf-net_2020} introduced a parallel partial decoder, called Inf-Net, which combines aggregation of high-level features to generate a global map. This is achieved through the use of convolutional hierarchies.

A U-Net-based model, named U-Net++, was applied to high-resolution CT images for Covid-19 detection in \cite{chen_deep_2020}. Furthermore, in \cite{ardakani_application_2020} a system for the detection of Covid-19 using 10 variants of CNNs in CT images is proposed, including AlexNet, VGG-16, VGG-19, SqueezeNet, GoogleNet, MobileNet-V2, ResNet-18, ResNet-50, ResNet-101, and Xception. ResNet-101 and Xception outperformed the remaining ones. AlexNet and Inception-V4 were also used for Covid-19 detection in CT scans in \cite{cifci_deep_2020}. The framework presented in \cite{singh_classification_2020} used a CNN and an Artificial Neural Network Fuzzy Inference System (ANNFIS) to detect Covid-19, whereas a Stack Hybrid Classification (SHC) scheme based on ensemble learning is proposed in \cite{farid_novel_2017}.
Focusing on segmentation, a type 2 fuzzy clustering system combined with a Super-pixel based Fuzzy Modified Flower Pollination Algorithm is proposed in \cite{chakraborty_sufmofpa_2021} for Covid-19 CT image segmentation.

Finally in \cite{katsamenis_transfer_2020} the experimental results indicate that the transfer learning approach outperforms the performance obtained without transfer learning, for the Covid-19 classification task in chest X-ray images, using deep classification models, such as convolutional neural networks (CNNs).

\section{Experimental setup}

\begin{figure*}[ht!]
     \centering
         \centering
         \includegraphics[width=\textwidth]{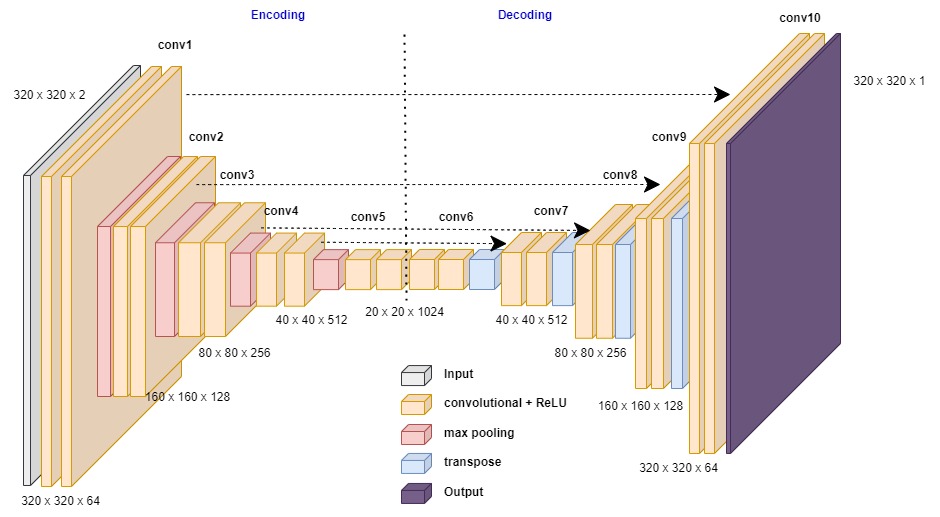}
         \caption{U-Net Architecture}
         \label{fig:U-Net Architecture}
\end{figure*}

The U-Net and the data transformation scripts, were developed in Python 3 using TensorFlow and Keras libraries. The models were trained in a VM using Unix MATE OS with 8 Core CPU and 64GB RAM provided GRNet Synefo service. Figure \ref{fig:U-Net Architecture} presents the architecture of the U-Net model.

\begin{table*}[hbt!]
\small
\caption{Datasets' properties}
\centering
\begin{tabular}{ccccccccc}%lllllllll ccccccccc

% \hline
\multicolumn{1}{c}{\textbf{Set Name}} &
  \multicolumn{1}{c}{\textbf{\begin{tabular}[c]{@{}c@{}}\# DICOM Files\end{tabular}}} &
  \multicolumn{1}{c}{\textbf{\# Slides}} &
  \multicolumn{1}{c}{\textbf{\begin{tabular}[c]{@{}c@{}}Slides with\\ lung areas\end{tabular}}} &
  \multicolumn{1}{c}{\textbf{\begin{tabular}[c]{@{}c@{}}Slides with \\ Covid areas\end{tabular}}} &
  \multicolumn{1}{c}{\textbf{\begin{tabular}[c]{@{}c@{}}Train set slides\\ (ratio 50\%/50\% \\ Covid/non-Covid slides)\end{tabular}}} &
  \multicolumn{1}{c}{\textbf{\begin{tabular}[c]{@{}c@{}}Validation set slides \\ (ratio 50\%/50\% \\ Covid/non-Covid slides)\end{tabular}}} &
  \multicolumn{1}{c}{\textbf{\begin{tabular}[c]{@{}c@{}}Test set \\Slides\end{tabular}}} &
  \multicolumn{1}{c}{\textbf{Epochs}} \\ \hline
CT 1-9 &
  9 &
  829 &
  713 &
  373 &
  440 &
  60 &
  214 &
  111 \\ \hline
CT 10 &
  10 &
  2581 &
  2156 &
  1351 &
  440 &
  60 &
  1656 &
  46 \\ \hline
  \label{table:1}
\end{tabular}
\end{table*}

\subsection{Datasets description}

To extract our results, we used two Lung Covid infected datasets, the Covid-19 CT segmentation dataset \cite{noauthor_Covid-19_nodate-1} from which we used only the Segmentation dataset nr. 2, this includes 9 DICOM files of continuous lung CT scans and the 20th April update \cite{jun_Covid-19_2020}  which contains another 20 labeled Covid-19 CT scans, from this we used only the 10 files marked as Coronacases and not the Redeopedia ones. 
The reason for this, is because all DICOM files we used, contained data in the Hounsfield scale \cite{PMID:31613501} the Radeopedia set of DICOM files, contained pixel values in the range of 0-255 therefore we could not use it since it did not follow our normalization procedure.
The first set of 9 DICOM files (we refer to this set as CT 1-9), contained 829 slides of CT, having dimensions of 630x630 pixels and includes already hand annotated lung and Covid masks for each slide of it. Similarly, the Coronacases dataset contained 10 CT scans (we refer to this set as CT 10) with 2581 slides in total having dimensions of 512x512 and includes also annotated masks by radiologists of the lung and Covid areas. Both sets, include continuous slides of complete lung CT scans of the same patient and not slides of different patients in each DICOM file. 
To construct our dataset we used only the slides that include lung areas, in order to achieve better results by reducing the extra information of slides without lung areas. But looking at Table \ref{table:1}, we see that the data in the CT 1-9 dataset use 214 train + 60 validation + 440 test slides total of 714 slides, but we see that the slides with lung areas are in total 713. This happens since one of the slides, marks a tiny area of few pixels as Covid, even if there is no lung or Covid in it, like in Figure \ref{fig:False Covid annotations} (b). This is due to a human error in the annotation procedure. 
For the normalization process, we resized the images to 320x320 pixels using Nearest Neighbor Interpolation and we kept only the Hounsfield values in the range of -970 to -150. All the information that is needed in our paradigm of Covid segmentation in lung areas, relies in this spectrum of Hounsfield scale. To achieve this, we normalize each pixel based on the following type

\begin{equation}
\frac{(pixelVal-minHounsfield)}{(maxHounsfield-minHounsfield)}=\frac{(pixelVal+970)}{(-150+970)}
\end{equation}

For every pixel value we get that is greater than 1. we assign the value 1 and every value less than 0. we assign the value 0. This way our dataset includes only values in the range [0.,1.]  
The radiologists have also marked in separate files the lung masks and the Covid masks for each slide of the CT scans we used. Therefore we arranged our Training Input in the form of n X 320 X 320 X 2 since in the first channel we used the normalized values of the CT slides and in the second channel we used the lung masks of each slide using binary values of 0,1 (1 in pixels that are marked as lung, 0 elsewhere). The output was having the form of n X 320 X 320 X 1 signaled with binary values 0,1 (1 in pixels that are marked as Covid, 0 elsewhere).

\subsection{Performance metrics}
In order to objectively evaluate our results, four different metrics are considered: accuracy, precision, recall, and the F1‐score, which is directly calculated from precision and recall values. Accuracy (ACC) is defined as: \begin{equation}
{ACC}=\frac{(TP+TN)}{(TP+TN+FP+FN)}
\end{equation}
where the nominator contains the true positives (TP) and true negatives (TN) samples, while denominator contains the TP and TN and false positives (FP) and false negatives (FN). Precision, recall and F1‐score are given as: 
\begin{equation}
{Pr}=\frac{TP}{(TP+FP)}
\end{equation}

\begin{equation}
{Re}=\frac{TP}{(TP+FN)}
\end{equation}

\begin{equation}
{F1}=\frac{Pr\times Re}{(Pr+Re)}
\end{equation}

\subsection{Experimental results}
With the pre-described formation of both datasets, we started the training of a Unet using the CT 1-9 data, with 4 encoding/decoding layers having as input 440 x 320 x 320 x 2 were in the 1\textsuperscript{st} channel we have the data of the CT scan and in the 2\textsuperscript{nd} channel we have the lung mask of the specific CT slide. This way the model focuses only in the lung areas of the CT scans in the learning process. In the output we have 440 x 320 x 320 x 1 where in the only channel we have the masks data of the Covid areas. We also used a validation set of 60 x 320 x 320 x 2 as input and 60 x 320 x 320 x 1 as output. In the Unet we used (i.e. Figure \ref{fig:U-Net Architecture}) the rectified linear activation unit (ReLU) function for the 3x3 conv layers and the Sigmoid activation in the 1x1 conv layer, to get output values in the range (0,1), a learning rate of 0.0001, a batch size of 45 and the shuffling enabled. The max epochs were 200 from which it only used 111 till the early stopping engaged. 

We then extracted metrics for the test data of the CT 1-9 set (i.e. 214 slides) these were F1 score, Accuracy, Precision and Recall (see Table \ref{table:PerfMetrics}), from which we got a great Accuracy value with an average of 0.9973 and an F1 score with an average of 0.7832 in total. 

\begin{table}[hbt!]
\caption{Performance Metrics}
\begin{tabular}{rcccc}
\cline{2-5}
\multicolumn{1}{l}{} & Acc & Pre & Rec   & F1       \\ \hline
CT 1-9               & 0.997375 & 0.751451  & 0.833533 & 0.783245 \\
CT 10 (no retrain)     & 0.995027 & 0.628606  & 0.647012 & 0.613689 \\
CT 10 (retrain)               & 0.997397 & 0.813515  & 0.863668 & 0.827899
 \label{table:PerfMetrics}
\end{tabular}
\end{table}

\begin{figure}[hbt!]
     \centering
     \begin{subfigure}[b]{0.3\columnwidth}
         \centering
         \includegraphics[width=25mm,height=25mm]{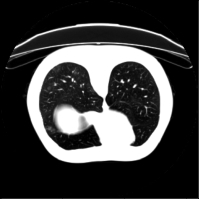}
        %  \caption{CT scan}
        %  \label{fig:CT scan}
     \end{subfigure}
     \hfill
     \begin{subfigure}[b]{0.3\columnwidth}
         \centering
         \includegraphics[width=25mm,height=25mm]{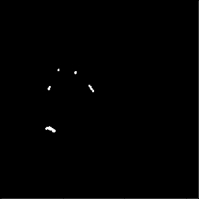}
        %  \caption{Ground Truth}
        %  \label{fig:Ground Truth with false annotations}
     \end{subfigure}
     \hfill
     \begin{subfigure}[b]{0.3\columnwidth}
         \centering
         \includegraphics[width=25mm,height=25mm]{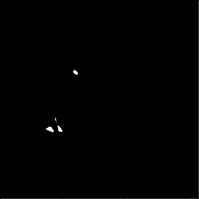}
        %  \caption{Prediction}
        %  \label{fig:Prediction}
     \end{subfigure}
        \caption{1\textsuperscript{st} Dataset Bad Prediction Example; CT scan (left), Ground Truth (middle), Prediction (right)}
        \label{fig:1st Dataset Bad Prediction Example}
\end{figure}

In Figure \ref{fig:1st Dataset Bad Prediction Example} we can see that the model performs great in the test set of the 1\textsuperscript{st} dataset, even in a bad scenario where the F1 Score is 0.4164 for this specific image. 
\begin{figure}[hbt!]
     \centering
     \begin{subfigure}[b]{0.3 \columnwidth}
         \centering
         \includegraphics[width=25mm,height=25mm]{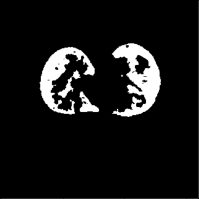}
        %  \caption{CT scan}
        %  \label{fig:CT scan}
     \end{subfigure}
     \hfill
     \begin{subfigure}[b]{0.3 \columnwidth }
         \centering
         \includegraphics[width=25mm,height=25mm]{samples/Figures_Covid/CT1-9_-_good_result_-_ground_truth_mask.jpg}
        %  \caption{Ground Truth}
        %  \label{fig:Ground Truth with false annotations}
     \end{subfigure}
     \hfill
     \begin{subfigure}[b]{0.3\columnwidth }
         \centering
         \includegraphics[width=25mm,height=25mm]{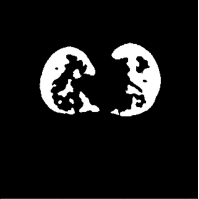}
        %  \caption{Prediction}
        %  \label{fig:Prediction}
     \end{subfigure}
        \caption{1\textsuperscript{st} Dataset Good Prediction Example; CT scan (left), Ground Truth (middle), Prediction (right)}
        \label{fig:1st Dataset Good Prediction Example}
\end{figure}

Afterwards we moved with the extraction of the same metrics, using all slides of set of the second dataset (CT 10). From these we got an average accuracy of 0.995 but an F1 score of 0.6137. Due to the low F1 score, we re-trained our model using a small portion of this dataset. From CD 10 dataset we had a similar size of train and validation sets (440 and 60) and we continued the train of the previous model but now we reduced the max epochs to 50. From these epochs the train process used 46 till the early stopping engaged. We again used the re-trained model in the test set of the 2\textsuperscript{nd} dataset which includes 1656 slides to obtain our metrics. 
\begin{figure}[hbt!]
     \centering
     \begin{subfigure}[b]{0.3\columnwidth}
         \centering
         \includegraphics[width=25mm,height=25mm]{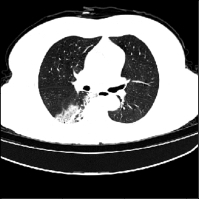}
        %  \caption{CT scan}
        %  \label{fig:CT scan}
     \end{subfigure}
     \hfill
     \begin{subfigure}[b]{0.3\columnwidth}
         \centering
         \includegraphics[width=25mm,height=25mm]{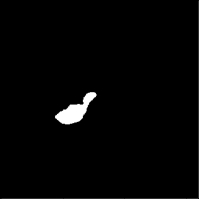}
        %  \caption{Ground Truth}
        %  \label{fig:Ground Truth with false annotations}
     \end{subfigure}
     \hfill
     \begin{subfigure}[b]{0.3\columnwidth}
         \centering
         \includegraphics[width=25mm,height=25mm]{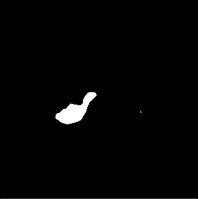}
        %  \caption{Prediction}
        %  \label{fig:Prediction}
     \end{subfigure}
        \caption{2\textsuperscript{nd} Dataset Good Prediction Example; CT scan (left), Ground Truth (middle), Prediction (right)}
        \label{fig:2nd Dataset Good Prediction Example}
\end{figure}

We must note here that all the slides that have been used from both datasets, had an area marked as lung in the accompanied mask files from the radiologists. If a slide did not include a lung area, it was not included in the any of the trained/validation or test set data. From the extraction of metrics in the retrained model, we got an average accuracy of 0.9974 and an average F1 score of 0.8279 which was an improvement compared to the 0.6137 F1 before the retrain of the model.

\begin{figure}[hbt!]
     \centering
     \begin{subfigure}[b]{0.3\columnwidth}
         \centering
         \includegraphics[width=25mm,height=25mm]{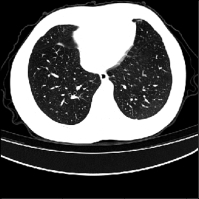}
        %  \caption{CT scan}
        %  \label{fig:CT scan}
     \end{subfigure}
     \hfill
     \begin{subfigure}[b]{0.3\columnwidth}
         \centering
         \includegraphics[width=25mm,height=25mm]{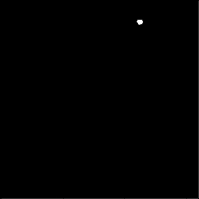}
        %  \caption{Ground Truth}
        %  \label{fig:Ground Truth with false annotations}
     \end{subfigure}
     \hfill
     \begin{subfigure}[b]{0.3\columnwidth}
         \centering
         \includegraphics[width=25mm,height=25mm]{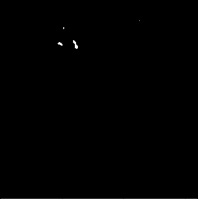}
        %  \caption{Prediction}
        %  \label{fig:Prediction}
     \end{subfigure}
        \caption{2\textsuperscript{nd} Dataset Bad Prediction Example; CT scan (left), Ground Truth (middle), Prediction (right)}
        \label{fig:2nd Dataset Bad Prediction Example}
\end{figure}

In Figure \ref{fig:2nd Dataset Bad Prediction Example} see an example of how the re-trained model performs, even in the bad scenario where the F1 Score of is 0.0136 we cannot clearly see an actual GGO on the CT image, even if in the Covid masks the radiologists marks one. 

\section{Dataset limitations}
\begin{figure}[hbt!]
     \centering
     \begin{subfigure}[b]{0.48\columnwidth}
         \centering
         \includegraphics[width=\columnwidth]{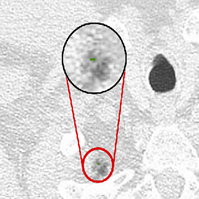}
        %  \caption{False annotated area 1}
        %  \label{fig: False annotated area 1}
     \end{subfigure}
     \hfill
     \begin{subfigure}[b]{0.48\columnwidth}
         \centering
         \includegraphics[width=\columnwidth]{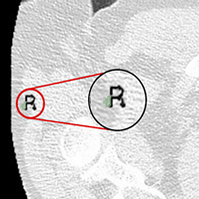}
        %  \caption{False annotated area 2}
        %  \label{fig:False annotated area 2; Are}
     \end{subfigure}
        \caption{False Covid annotations; Area 1 (left), Area 2 (right)}
        \label{fig:False Covid annotations}
\end{figure}

In the datasets we have used, we spotted some inconsistencies, since its annotation was made by hand. Specifically in our pre-train analysis, we found that in DICOM files there were slides with a false marked Covid areas. As we can see in Figure \ref{fig:False Covid annotations} (b) the radiologist has a marked Covid area in a section that there is no lung at all.

\begin{figure}[hbt!]
     \centering
     \begin{subfigure}[b]{0.3\columnwidth}
         \centering
         \includegraphics[width=25mm,height=25mm]{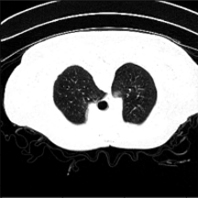}
        %  \caption{CT scan}
        %  \label{fig:CT scan}
     \end{subfigure}
     \hfill
     \begin{subfigure}[b]{0.3\columnwidth}
         \centering
         \includegraphics[width=25mm,height=25mm]{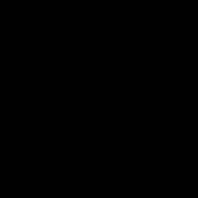}
        %  \caption{Ground Truth}
        %  \label{fig:Ground Truth with no annotations}
     \end{subfigure}
     \hfill
     \begin{subfigure}[b]{0.3\columnwidth}
         \centering
         \includegraphics[width=25mm,height=25mm]{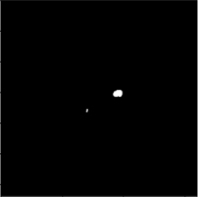}
        %  \caption{Prediction}
        %  \label{fig:Prediction}
     \end{subfigure}
        \caption{Missing annotated Covid areas, in the original dataset; CT scan (left), Ground Truth (middle), Prediction (right)}
        \label{fig:Not annotated Covid areas}
\end{figure}

We have also spotted a case where our model predicted a Covid area even if it was not marked as one by the radiologist. From the Figure \ref{fig:Not annotated Covid areas} we can see that there is a GGO in the CT scan area, predicted by our model, even if there is no mask in ground truth.

Taking account the previously mentioned paradigms, we cannot be sure of the extent of these faulty annotated areas, since the only chance to find them is to re-evaluate all annotations by a different radiologist and compare the results. 

\section{3D representations}
To assist the medical personnel which works with tools like Computed Tomography scans, we constructed a python script that exports a 3D representation of the CT scan data and the Covid segments produced by our model. 
At first, we extracted for every slide in a CT the color value of each pixel. We created an array of color values only in the lung areas, each entry of the array was having the form of (x position, y position, z position, color value) The z position was computed starting from 0 and adding a constant value for every new slide that we export its data. We did the same procedure for the Covid masks that our model predicted and for the ground truth, but in these the color value was in binary (i.e. 0 or 1). We then exported these arrays in comma separated files which we then imported to the ParaView visualizer \cite{noauthor_paraview_nodate} using the table to point filtering. 

\begin{figure}[hbt!]
     \centering
     \begin{subfigure}[b]{0.3\columnwidth}
         \centering
         \includegraphics[width=\columnwidth]{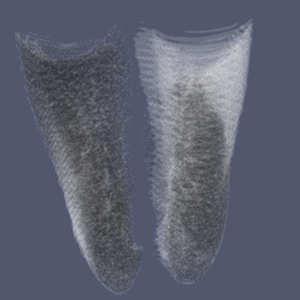}
        %  \caption{CT scan}
        %  \label{fig:CT scan}
     \end{subfigure}
     \hfill
     \begin{subfigure}[b]{0.3\columnwidth}
         \centering
         \includegraphics[width=\columnwidth]{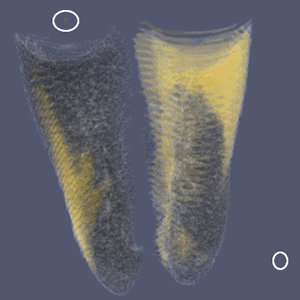}
        %  \caption{Ground Truth}
        %  \label{fig:Ground Truth with false annotations}
     \end{subfigure}
     \hfill
     \begin{subfigure}[b]{0.3\columnwidth}
         \centering
         \includegraphics[width=\columnwidth]{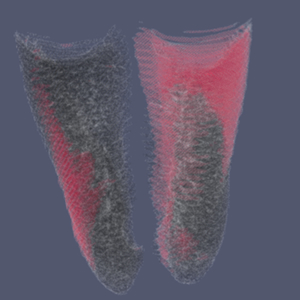}
        %  \caption{Prediction}
        %  \label{fig:Prediction}
     \end{subfigure}
        \caption{3D Representation; CT scan (left), Ground Truth (middle), Prediction (right)}
        \label{fig:3D Representation}
\end{figure}

With the 3D reconstruction, the evaluation of a patience status is easier for the medical personnel, as they can see the whole lung and not evaluate separate CT slides which can be significant in number. With this tool it is also easier to estimate how a treatment performs in time, using different 3D representations of the same patient. 
From the Fig \ref{fig:3D Representation} we can compare the model prediction with the Ground Truth but we can also see that in the Ground Truth, there are faulty annotated Covid areas outside the lung segments (marked in the Ground Truth), made by the radiologists. 

\begin{acks}
This research has been co‐financed by European Union’s Horizon 2020 research and innovation programme under grant agreement No 883441 for the STAMINA Innovation action.
\end{acks}

%%
%% The next two lines define the bibliography style to be used, and
%% the bibliography file.
% \bibliographystyle{ACM-Reference-Format}
% \bibliography{Evaluating_DL_Transferability_for_Covid_3D_Localization_Using_CT_scans}

\printbibliography
\end{document}